\documentclass[letterpaper,twoside,english,prl, showpacs,twocolumn]{revtex4}
\usepackage{times}
\usepackage[T1]{fontenc}
\usepackage{amsmath}
\usepackage{graphicx}
\usepackage{amssymb}

\makeatletter
\usepackage{babel}
\makeatother
\begin{document}

\title{Long Chaotic Transients in Complex Networks }

\author{Alexander Zumdieck$^{1}$%
\footnote{present address: Max-Planck-Institut f\"ur Physik komplexer Systeme,
01187 Dresden, Germany%
}}

\author{Marc Timme$^{1}$}

\author{Theo Geisel$^{1,2}$}

\author{Fred Wolf$^{1,2}$}

\affiliation{$^{1}$Max-Planck-Institut f\"ur Str\"omungsforschung and}

\affiliation{Fakultät für Physik, Universität Göttingen, 37073 G\"ottingen, Germany}

\affiliation{$^{2}$Kavli Institute for Theoretical Physics, University of California
Santa Barbara, California 93106}

\begin{abstract}
We show that long chaotic transients dominate the dynamics of randomly
diluted networks of pulse-coupled oscillators. This contrasts with
the rapid convergence towards limit cycle attractors found in networks
of globally coupled units. The lengths of the transients strongly
depend on the network connectivity and varies by several orders of
magnitude, with maximum transient lengths at intermediate connectivities.
The dynamics of the transient exhibits a novel form of robust synchronization.
An approximation to the largest Lyapunov exponent characterizing the
chaotic nature of the transient dynamics is calculated analytically.
\end{abstract}


\maketitle
The dynamics of complex networks~\cite{nw_struct} is a challenging
research topic in physics, technology and the life sciences. Paradigmatic
models of units interacting on networks are pulse- and phase-coupled
oscillators~\cite{strog:01}. Often attractors of the network dynamics
in such systems are states of collective synchrony~\cite{ernst:95,gerst:96,mirollo:90,pesk:84,twg:02a,twg:02b,sync,bressloff,hansel,herz}.
Motivated by synchronization phenomena observed in biological systems,
such as the heart~\cite{pesk:84} or the brain~\cite{sing:89},
many studies have investigated how simple pulse-coupled model units
can synchronize their activity. Here a key question is whether and
how rapid synchronization can be achieved in large networks. It has
been shown that fully connected networks as well as arbitrary networks
of non-leaky integrators can synchronize very rapidly~\cite{mirollo:90,gerst:96}.
Biological networks however, are typically composed of dissipative
elements and exhibit a complicated connectivity.

In this Letter, we investigate the influence of diluted network connectivity
and dissipation on the collective dynamics of pulse-coupled oscillators.
Intriguingly, we find that the dynamics is completely different from
that of globally coupled networks or networks of non-leaky units,
even for moderate dissipation and dilution: long chaotic transients
dominate the network dynamics for a wide range of connectivities,
rendering the attractors (simple limit cycles) irrelevant. Whereas
the transient length is shortest for very high and very low connectivity,
it becomes very large for networks of intermediate connectivity. The
transient dynamics exhibits a robust form of synchrony that differs
strongly from the synchronous dynamics on the limit cycle attractors.
We quantify the chaotic nature of the transient dynamics by analytically
calculating an approximation to the largest Lyapunov exponent on the
transient.

We consider a system of $N$ oscillators~\cite{mirollo:90,ernst:95}
that interact on a directed graph by sending and receiving pulses.
For concreteness we consider asymmetric random networks in which every
oscillator $i$ is connected to an other oscillator $j\neq i$ by
a directed link with probability $p$. A phase variable $\phi_{j}(t)\in[0,1]$
specifies the state of each oscillator $j$ at time $t$. In the absence
of interactions the dynamics of an oscillator $j$ is given by\begin{equation}
\textrm{d}\phi_{j}(t)/\textrm{d}t=1\,.\label{eq:timeevol}\end{equation}
 When an oscillator $j$ reaches the threshold, $\phi_{j}(t)=1$,
its phase is reset to zero, $\phi_{j}(t^{+})=0$, and the oscillator
emits a pulse that is sent to all oscillators $i$ possessing an in-link
from $j$. After a delay time $\tau$ this pulse induces a phase jump
in the receiving oscillator $i$ according to\begin{equation}
\phi_{i}((t+\tau)^{+}):=\min\{ U^{-1}(U(\phi_{i}(t+\tau))+\varepsilon_{ij}),1\}\label{eq:phasjump}\end{equation}
 which depends on its instantaneous phase $\phi_{i}(t+\tau)$, the
excitatory coupling strength $\varepsilon_{ij}\ge0$, and on whether
the input is sub- or supra-threshold. The phase dependence is determined
by a twice continuously differentiable function $U(\phi)$ that is
assumed to be strictly increasing, $U'(\phi)>0$, concave (down),
$U''(\phi)<0$, and normalized such that $U(0)=0$ and $U(1)=1$ (cf.~\cite{mirollo:90,TWG:03a}). 

This model, originally introduced by Mirollo and Strogatz~\cite{mirollo:90},
is equivalent to different well known models of interacting threshold
elements if $U(\phi)$ is chosen appropriately (cf.~\cite{TWG:03a}).
The results presented in this Letter are obtained for $U_{b}(\phi)=b^{-1}\ln(1+(e^{b}-1)\phi)$,
where $b>0$ parameterizes the curvature of $U$, that determines
the strength of the dissipation of individual oscillators. The function
$U$ approaches the linear, non-leaky case in the limit $\lim_{b\rightarrow0}U_{b}(\phi)=\phi$.
Other nonlinear choices of $U\neq U_{b}$ give results similar to
those reported below. The considered graphs are strongly connected,
i.e. there exists a directed path from any node to any other node.
We normalize the total input to each node $\sum_{j=1}^{N}\varepsilon_{ij}=\varepsilon$
such that the fully synchronous state ($\phi_{i\ }(t)\equiv\phi_{0}(t)$
for all $i$) exists~\cite{twg:02b}. Furthermore for any node $i$
all its $k_{i}$ incoming links have the same strength $\varepsilon_{ij}=\varepsilon/k_{i}$~.

\begin{figure*}
\begin{center}\includegraphics[%
  width=0.90\textwidth,
  keepaspectratio]{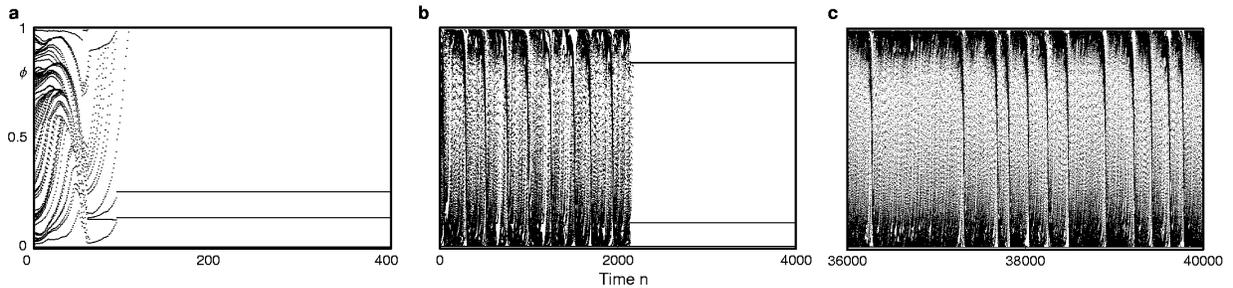}\end{center}

\caption{\label{fig:dynamics}Long chaotic transients towards periodic attractor
states in complex networks. The panels show the dynamics of random
networks of $N=100$ oscillators ($b=1.0,$~$\varepsilon=0.1$,~$\tau=0.1$).
The phases of all oscillators are marked along the vertical axis just
after an arbitrary but fixed reference oscillator is reset. Time is
thus measured by the number of resets $n$ of the reference oscillator.
The length of the transients increases quickly with dilution: (a)
for the fully connected ($p=1$) network the transient length is $T\approx10^{2}$,
whereas (b) $T\approx2\cdot10^{3}$ for $p=0.97$ and (c) no attractor
reached up to $n=10^{5}$ ($T>10^{5}$) for $p=0.80$. }
\end{figure*}
Numerical investigations of all-to-all coupled networks ($p=1$) show
rapid convergence from arbitrary initial conditions to periodic orbit
attractors (cf.~\cite{mirollo:90,twg:02a}), in which several synchronized
groups of oscillators (clusters) coexist~\cite{ernst:95,twg:02a}.
In general we find that the transient length $T$, i.e. the time the
system needs to reach an attractor, is short for all-to-all coupled
networks and depends only weakly on network size, for instance $T\approx10^{2}$
for $N=100$ oscillators {[}see e.g. Fig.~\ref{fig:dynamics}(a){]}. 

\begin{figure}
\begin{center}\includegraphics[%
  width=0.70\columnwidth,
  keepaspectratio]{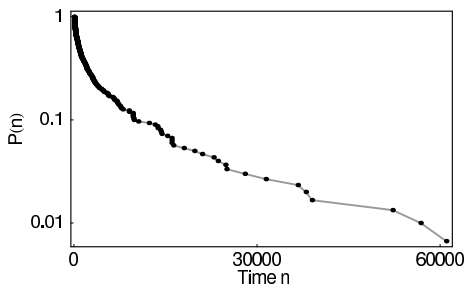}\end{center}

\caption{\label{fig:pvont} Typical transients in complex networks are very
long. Distribution of transient lengths on randomly diluted networks
($N=16$, $p=0.8$, $b=3$, $\varepsilon=0.1,\,\tau=0.1$, histogram
of 300 random networks). }
\end{figure}

In contrast to fully connected networks, diluted networks exhibit
largely increased transient times: eliminating just $3\%$ of the
links ($p=0.97$) leads to an increase in $T$ of one order of magnitude
{[}$T\approx10^{3}$, Fig.~\ref{fig:dynamics}(b){]}. The system
finally settles on an attractor which is similar to the one found
in the fully connected network, i.e. a periodic orbit with several
synchronized clusters. Further dilution of the network causes the
transient length to grow extremely large {[}$T>10^{5}$, Fig.~\ref{fig:dynamics}(c){]}.
The lifetime of an individual transient typically depends strongly
on the initial condition. We observed the dynamics started from many
randomly chosen initial phase vectors distributed uniformly in $[0,1]^{N}$
and typically find a wide range of transient times (Fig.~\ref{fig:pvont}).
\begin{figure}
\begin{center}\includegraphics[%
  width=0.80\columnwidth,
  keepaspectratio]{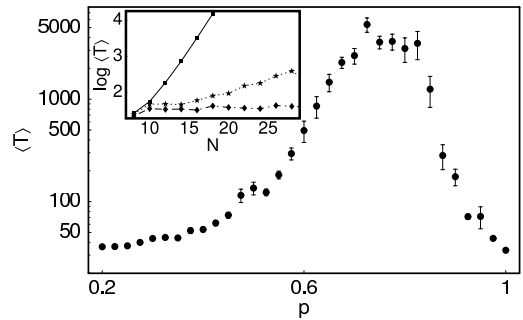}\end{center}

\caption{\label{fig:tvsp}The average transient length $\langle T\rangle$
depends non-monotonically on the network connectivity $p$. The longest
transients are found for intermediate connectivities. Average over
100 initial conditions ($N=16$, $b=3,\varepsilon=0.1,\,\tau=0.1$).
Error bars are standard error of mean. The inset shows the rapid growth
of $\langle T\rangle$ with network size $N$ (boxes: $p=0.8$, stars:
$p=0.95$, diamonds: $p=1.0$ ). }
\end{figure}

We systematically studied the average transient length in dependence
on the average connectivity $p$. Surprisingly the average transient
length, a dynamical feature of the network, depends non-monotonically
on the network connectivity $p$ (Fig.~\ref{fig:tvsp} ), whereas
many structural properties of random graphs such as the average path
length between two vertices are monotonic in $p$ \cite{bolobas}.
The average transient length is short for low and high connectivity
values $p$, but becomes very large for intermediate connectivities,
even for only weakly diluted networks. Moreover we find that the mean
lifetime $\langle T\rangle$ grows exponentially with the network
size $N$ for diluted networks whereas it is almost independent of
network size for fully-connected networks (inset of Fig.~\ref{fig:tvsp}).
This renders long transients the dominant form of dynamics for all
but strongly diluted or fully connected networks. The transient length
defines a new, collective time scale that is much larger than the
natural period, $1$, of an individual oscillator and the delay time,
$\tau$, of the interactions. This separation of time scales makes
it possible to statistically characterize the dynamics on the transient
(cf. \cite{ttel:91}). 

\begin{figure}
\begin{center}\includegraphics[%
  width=0.70\columnwidth,
  keepaspectratio]{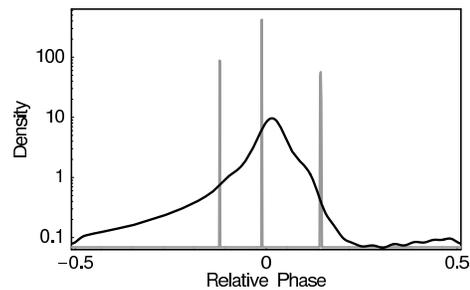}\end{center}

\caption{\label{fig:PhasDens}The oscillators form a single, slightly dispersed
cluster during the transient (black) and several precisely synchronized
clusters on the attractor (grey). The figure shows the density of
phases of the oscillators relative to their average phase. Data shown
for the dynamics displayed in Fig.~\ref{fig:dynamics}(b) averaged
over $n=500\dots1500$ (transient) and $n=2500\dots3500$ (attractor). }
\end{figure}

What are the main features of the transient dynamics? We determined
the statistical distribution of phases of the oscillators during the
transient. Interestingly the oscillators exhibit a novel kind of synchrony
during the transient: most of the units form a single, roughly synchronized
cluster (Fig. \ref{fig:PhasDens}). This cluster is robust in the
sense that it always contains approximately the same number of oscillators
with the same spread of relative phases although it continuously absorbs
and emits oscillators. On the attractor, however, the oscillators
are organized in several precisely synchronized clusters (cf. Fig.
\ref{fig:PhasDens}), such that the transient dynamics is completely
different from the dynamics on the attractor. 

\begin{figure}
\begin{center}\includegraphics[%
  width=0.70\columnwidth,
  keepaspectratio]{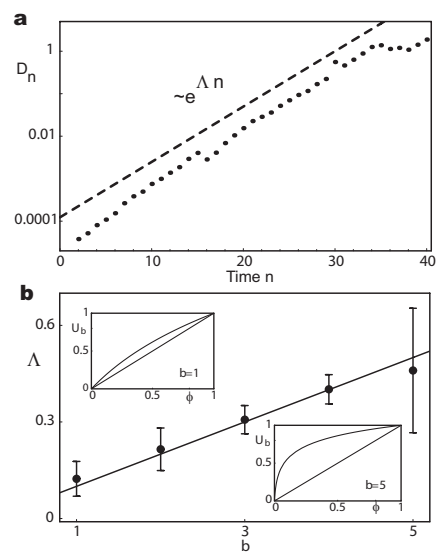}\end{center}

\caption{\label{fig:Pivsb}(a) The distance $D_{n}$ between two nearby trajectories
grows exponentially with time $n$. The slope of the dashed line quantifies
the rate of divergence measured by the largest Lyapunov exponent $\Lambda$.
The jumps at $n=16$ and $n=29$ are due to the choice of the reference
oscillator. (b) The largest Lyapunov exponent $\Lambda$ depends linearly
on the curvature $b$ of the interaction function $U_{b}$. These
numerical results (disks with error bars) agree with the analytical
prediction $\Lambda=b\cdot\varepsilon$ (continuous line) \cite{bdeviation}.
Error bars denote the standard deviation for 100 random networks ($N=100$,
$p=0.75$, $\varepsilon=0.1,\,\tau=0.1$). Inset: interaction function
$U_{b}(\phi)$ for $b=1$ and $b=5$.}
\end{figure}

Numerically we find that nearby trajectories diverge exponentially
with time, an indication of chaos {[}Fig.~\ref{fig:Pivsb}(a){]}.
To further quantify the chaotic nature of the transient dynamics,
we determined the speed of divergence of two nearby trajectories both
by numerical measurements and by analytically estimating the largest
Lyapunov exponent on the transient. Let \{$t_{n}$:~$n\in\mathbb{N}_{0}$\}
be a set of firing times of an arbitrary, fixed reference oscillator
and $\phi_{i}^{1}(t_{n}),\phi_{i}^{2}(t_{n})$, $i\in\{1,\dots,N\}$,
the set of phases of all oscillators $i$ of two transient trajectories
1 and 2. We then define the distance \begin{equation}
D_{n}:=\sum_{i=1}^{N}|\phi_{i}^{1}(t_{n})-\phi_{i}^{2}(t_{n})|_{c}\label{eq:ddef}\end{equation}

between these two trajectories at time $t_{n}$ where $|\cdot|_{c}$
denotes the distance of two points on a circle with circumference
1. For a small initial separation $D_{0}\ll1$ at time $t_{0}$ the
distance $D_{n}$ at later time $t_{n}$ scales like {[}Fig. \ref{fig:Pivsb}(a){]}\begin{equation}
D_{n}\approx D_{0}e^{\Lambda n}\,,\label{eq:expdiv}\end{equation}
 quantifying the speed of divergence by the largest Lyapunov exponent
$\Lambda$. To analytically estimate $\Lambda$, we determine the
phase advance of oscillator $i$ evoked by a single pulse received
from oscillator $j$,\begin{equation}
\phi_{i}(t^{+})=a_{ij}\phi_{i}(t)+c_{ij}\,,\label{eq:phasadv}\end{equation}
 where we have used the definition of $U_{b}(\phi)$ in Eq.~(\ref{eq:phasjump})
and considered only sub-threshold input. We obtain \begin{equation}
a_{ij}:=\exp(b\varepsilon_{ij})\ge1\,,\label{eq:aij}\end{equation}

independent of the delay $\tau$ and the constants $c_{ij}$ which
are independent of $\phi_{j}(t)$. The magnitude by which a single
pulse increases the difference $|\phi_{i}^{1}(t^{+})-\phi_{i}^{2}(t^{+})|_{c}=a_{ij}|\phi_{i}^{1}(t)-\phi_{i}^{2}(t)|_{c}$
is thus only determined by $a_{ij}$. If an oscillator $i$ receives
exactly one pulse from all its upstream (i.e. presynaptic) oscillators
between $t_{n}$ and $t_{n+1}$, the total increase due to all these
pulses is determined by\begin{equation}
A_{i}:=\prod_{j=1}^{N}a_{ij}=\exp\left(b\sum_{j=1}^{N}\varepsilon_{ij}\right).\label{eq:advall}\end{equation}

The normalization $\sum_{j=1}^{N}\varepsilon_{ij}=\varepsilon$ implies
$A_{i}=\exp(b\varepsilon)=:A$ for all oscillators $i$. Noting that
the network is in an asynchronous state during the transient and assuming
that between $t_{n}$ and $t_{n+1}$ each oscillator fires exactly
once we find $D_{n+1}\approx AD_{n}$. The largest Lyapunov is then
approximated by\begin{equation}
\Lambda\approx\ln A=b\cdot\varepsilon.\label{eq:lypexp}\end{equation}

Numerical results are in good agreement with Eq.~(\ref{eq:lypexp})
for intermediate values of $b$ {[}\cite{bdeviation}, see Fig.~\ref{fig:Pivsb}(b){]}.
Note that there is no free parameter in Eq.~(\ref{eq:lypexp}). This
calculation shows that the curvature $b$ of $U_{b}$ and the coupling
strength $\varepsilon$ strongly influence the dynamics during the
transient by determining the largest Lyapunov exponent. To further
characterize the dynamics we investigated return maps of inter spike
intervals (i.e.~the time between two consecutive resets of individual
oscillators). This analysis revealed broken tori in phase space (not
shown). Moreover we find that the correlation dimension~\cite{grassberger}
of the transient orbit is low (e.g. $D_{\text{{corr.}}}\approx5$
for $N=100,$ $p=0.96$ and $b=3,\,\varepsilon=0.1,\,\tau=0.1$).
This indicates that the chaotic motion during the transient takes
place in the vicinity of quasiperiodic motion on a low dimensional
toroidal manyfold in phase space. Interestingly quasiperiodic motion
was also found in fully-connected networks of similar integrate-and-fire
oscillators \cite{vrees:96a}. 

How do the long chaotic transients come into existence? As a first
step towards answering this question we investigated in which region
of parameter space long chaotic transients occur. It turned out that
they do not sensitively depend on parameters but prevail for a substantial
region of parameter space in randomly diluted networks. The region
covered by long transients in parameter space (not shown) is similar
to that covered by unstable attractors for fully connected networks
(cf. \cite{twg:02a}). As, in general, long chaotic transients often
occur in situations like crises or attractor destructions \cite{tel:94},
we conjecture that long transients occur in this system because diluting
the network converts unstable attractors that prevail in phase space
into unstable (presumably non-attracting) periodic orbits, inducing
sensitive dependence on initial conditions and thus chaotic transient
dynamics. The interesting open question about the dynamical origin
of the transient needs further investigation.

In summary, we described a novel type of dynamics in complex networks
of pulse-coupled oscillators: long chaotic transients. These transients
dominate the dynamics for a wide range of parameters and become prevalent
for large networks, thus rendering the dynamics on the attractors
irrelevant to the observed behavior. This is in stark contrast to
the rapid convergence found in fully connected networks as well as
in networks of non-leaky elements. The transient length defines a
new, collective timescale that is not present in the single unit dynamics.
Interestingly it is maximal for intermediate connectivities in contrast
to many structural network properties. The transient dynamics exhibits
a rapid and robust form of synchronization: the oscillators form a
roughly synchronized cluster. For the transient dynamics we approximately
calculated the largest Lyapunov exponent analytically; this is rarely
possible for any high dimensional system. The approximation is in
good quantitative agreement with the exponential divergence of nearby
trajectories found in numerical simulations. 

Previous studies on the synchronization dynamics of pulse-coupled
oscillators have focused on globally coupled networks or on systems
of individual elements without dissipation (see e.g. \cite{ernst:95,gerst:96}).
Our results show that the combination of dissipation and complex connectivity
can create a qualitatively distinct network dynamics. More generally,
our results emphasize that the network's structure can have a major
impact on its dynamics, as small structural changes induce fundamentally
different forms of behavior. This is very likely to occur not only
in networks of coupled oscillators but in many other complex networks,
too.

This research was supported in part by the National Science Foundation
under Grant No. PHY99-07949.


\begin{thebibliography}{13}
\bibitem{bdeviation}For values of $b$ outside of the range shown in Fig. \ref{fig:Pivsb}
deviations exist because not all oscillators fire exactly once during
one cycle. For small $b<1$ we observe irregular dynamics that is
markedly different from that shown in Fig.~\ref{fig:dynamics}; we
find in particular that individual oscillators remain in the dynamic
cluster for a long time which compromises the approximation of an
asynchronous state during the transient. For large $b>5$ the bent
character of $U_{b}$ gets more and more kink-like. Thus more pulses
are supra-threshold, the dynamics is strongly influenced by phase
resets and thus not adequately described by Eq.~(\ref{eq:phasadv}).
\end{thebibliography}
\end{document}